\documentclass[]{article}
\pdfoutput=1
\usepackage{spconf,amsmath,graphicx}
\usepackage{amsmath,graphicx}
\usepackage{amsfonts}
\usepackage{color}
\usepackage{caption}
\usepackage{subfigure}
\usepackage{tabularx}
\usepackage{amsthm}

\newtheorem{thm}{Theorem}

\newtheorem{lem}[thm]{Lemma}
\newtheorem{prop}[thm]{Proposition}
\newtheorem{defn}[thm]{Definition}

\title{Sampling of surfaces and functions in high dimensional spaces}
\twoauthors
  {Qing Zou}
	{Department of Mathematics, \\
           University of Iowa, IA, USA}
 {Mathews Jacob\thanks{This  work  is  supported  by NIH 1R01EB019961-01A1.}}%
	{Department of Electrical and Computer Engineering, \\
           University of Iowa, IA, USA}
\begin{document}
%\ninept
%
\maketitle
\begin{abstract}
We introduce a sampling theoretic framework for the recovery of smooth surfaces and functions living on smooth surfaces from few samples. The proposed approach can be thought of as a nonlinear generalization of union of subspace models widely used in signal processing. This scheme relies on an exponential lifting of the original data points to feature space, where the features live on union of subspaces. The low-rank property of the features are used to recover the surfaces as well as to determine the number of measurements needed to recover the surface. The low-rank property of the features also provides an efficient approach which resembles a neural network for the local representation of multidimensional functions on the surface; the significantly reduced number of parameters make the computational structure attractive for learning inference from limited labeled training data.
\end{abstract}

\vspace{-1em}
\section{Introduction}
\vspace{-1em}
Machine learning algorithms often exploit the extensive structure present in natural datasets, for visualization or to learn inference. For example, manifold embedding methods model data as points on simpler objects such as smooth curves or surfaces/manifolds in high dimensional spaces for visualization \cite{roweis2000nonlinear}. In practice, the measured real world data is often scarce, corrupted by extensive noise, missing data, and other measurement errors.  Thus, the recovery of noise-free data and/or learning of inference on the data from few noisy measurements are key problems in machine learning applications.  

The main focus of this work is to introduce a sampling theory for \textbf{(a)} recovery of high dimensional surfaces, and \textbf{(b)} local representation of functions that live on surfaces, from few measurements. In this work, we model the surface to be the zero set of a multidimensional band-limited function. As expected, a more band-limited function will translate to a smoother surface/curve; the bandwidth of the function serves as a measure of the regularity or complexity of the curve. Under this assumption, we show that the nonlinear features of any arbitrary point on such a curve, obtained by lifting of the points using an exponential map can be annihilated by the inner product with the coefficients of the level-set function. We show that the feature matrix is low-rank, which is  used to estimate the surface from few of its samples. We introduce sampling conditions that will guarantee the recovery of the surface with high probability, when the surface is irreducible or when it is the union of several irreducible surfaces. 

We generalize the above results to the local representation and recovery of band-limited multidimensional functions as the interpolation of a few samples on the surface. We also introduce sampling conditions that guarantee the perfect recovery of the function on the surface.  Specifically, the low-rank nature of the exponential features of the surface provides an elegant approach to locally represent the function using considerably lower number of parameters. This significant reduction in the number of free parameters offered by this local representation makes the learning of the function from finite samples tractable. We note that the computational structure of the representation is essentially a two layer kernel network. Note that the approximation is highly local; the true function and the local representation match only on the curve/surface, while they may deviate significantly on points not on the curve/surface. This behavior of the network may explain the sensitivity of practical machine learning algorithms to adversarial attacks. Specifically, the function approximation can be exact as long as the inputs are constrained to the data manifold; an adversarial attack designed to move the input away from the surface can result in unexpected function values.

This work builds upon our prior work \cite{poddar2018recovery,poddaricassp,poddar2018sampling}, where we considered the recovery of planar curves from several of the samples. This work in this paper extends the above results in three important ways \textbf{(i).} The planar results are generalized to the high dimensional setting in this work. \textbf{(ii).} The worst case sampling conditions are replaced by high-probability results, which are far less conservative, and are in good agreement with experimental results.  \textbf{(iii).} The sampling results are extended to the local representation of functions in this work.

 In particular, we show that the function can be evaluated as the interpolation of the function values on admissible anchor points on the curve/surface by a Dirichlet kernel function. 

\vspace{-2em}
\section{Background}
\vspace{-1em}
In this work, we model the surface as the zero level-set
\begin{equation}\label{leveleset}
\mathcal{S}=\{\mathbf{x}\in\mathbb{R}^n : \psi(\mathbf{x})=\mathbf {0}\}
\end{equation}
of the bandlimited function $\psi$:
\begin{equation}\label{tri}
\psi(\mathbf {x})=\sum_{\mathbf {k}\in\Lambda}\mathbf {c_k}\exp(j2\pi \mathbf {k}^T\mathbf {x}),\quad \mathbf {x}\in[0,1)^n.
\end{equation}
The cardinality of the set $\Lambda$, which is denoted by $|\Lambda|$ is the number of free parameters in the surface representation.  Note that the complexity of $\mathcal S$ grows with the bandwidth of $\psi$; 
$|\Lambda|$ is hence a measure of the complexity of the surface.  We denote the $\psi$ satisfying \eqref{leveleset}, whose  coefficient set $\{\mathbf{c_k} : \mathbf{k}\in\Lambda\}$ has the minimal support , as the minimal polynomial.

We now consider an arbitrary point $\mathbf x$ on the surface $\mathcal S$. By \eqref{leveleset}, we have $\psi(\mathbf x)=0$. Using the bandlimited representation in \eqref{tri}, we have
\begin{equation}\label{annihilation}
\sum_{\mathbf {k}\in\Lambda}\mathbf {c_k}\exp(j2\pi \mathbf {k}^T\mathbf {x})=\mathbf{c}^T\phi_{\Lambda}(\mathbf{x})=0.
\end{equation}
Here,  $\phi_{\Lambda}: \mathbb{R}^n\to\mathbb{C}^{|\Lambda|}$ is a nonlinear feature map, which lifts a point $\mathbf {x} \in [0,1)^n$ to a higher dimensional space:
\[\phi_{\Lambda}(\mathbf{x})=[\exp(j2\pi \mathbf {k}_1^T\mathbf {x}) \quad \cdots\quad \exp(j2\pi \mathbf {k}_{|\Lambda|}^T\mathbf {x})]^T.\]
Using the one-to-one correspondence between the trigonometric polynomials in \eqref{tri} and complex polynomials established by the one in \cite{poddar2018sampling,ongie2016off}, we define the irreducibility of the trigonometric polynomials. Specifically, we say that the trigonometric polynomial $\eta(\mathbf {x})$ is irreducible if the corresponding complex polynomial $\mathcal{P}[\eta]$ is an irreducible polynomial in $\mathbb{C}[z_1,\cdots,z_n]$.

\vspace{-0.6em}
\begin{defn}
A surface is termed as irreducible, if it is the zero set of an irreducible trigonometric polynomial. 
\end{defn}
\vspace{-1.5em}
\begin{lem}
The zero set of an irreducible minimal trigonometric polynomial can only have one connected component.
\end{lem}

\vspace{-2em}
\section{Surface recovery from samples}
\vspace{-1em}
Assume that we have $N$ samples on the curve. We define the 
the feature matrix of the sampling set $\mathbf{X}=\{\mathbf{x}_1, \cdots, \mathbf{x}_N\}$ as
$\Phi_{\Lambda}(\mathbf{X})=[\phi_{\Lambda}(\mathbf{x}_1) \quad\cdots\quad \phi_{\Lambda}(\mathbf{x}_N)]$. Since all of these points satisfy \eqref{leveleset}, we have 
\begin{equation}\label{nullspacecond}
\mathbf{c}^T~\Phi_{\Lambda}(\mathbf{X})=0.
\end{equation}
One can use the above null space relation to recover the coefficient vector $\mathbf c$ from the samples $\mathbf X$. Note that there is a one-to-one correspondence (up to scaling) between the coefficient set and the curve. Hence, to uniquely determine the zero level-set of $\psi(\mathbf{x})$, we require 
\begin{equation}\label{rankcond}
\mathrm{rank}(\Phi_{\Lambda}(\mathbf{X}))=|\Lambda|-1.
\end{equation}
The following result tells us when the feature matrix satisfy this rank condition, and thus guarantee the recovery of $\mathcal S$.

\vspace{-1.4em}
\subsection{Irreducible surfaces}
\vspace{-0.7em}
We first focus on the recover case, where $\mathcal S$ is an irreducible surface. These results generalize the recovery of subspaces or low-rank matrices from few samples. 

\begin{thm}\label{rankprop}
	Let $\mathcal S$ be an irreducible surface, which is the zero level-set of an irreducible trigonometric polynomial $\psi(\mathbf{x})$ whose bandwidth is given by $\Lambda$. Let $\{\mathbf{x}_1,\cdots, \mathbf{x}_N\} \in \mathcal S$ be $N$ samples, drawn randomly in an independently fashion. Then, the feature matrix $\Phi_{\Lambda}(\mathbf X)$ satisfies \eqref{rankcond} with high probability, provided $N\ge |\Lambda|-1$. 
\end{thm}

\vspace{-1.8em}
\begin{figure}[h!]
	\centering
	\subfigure[Curve]{\includegraphics[width=0.12\textwidth]{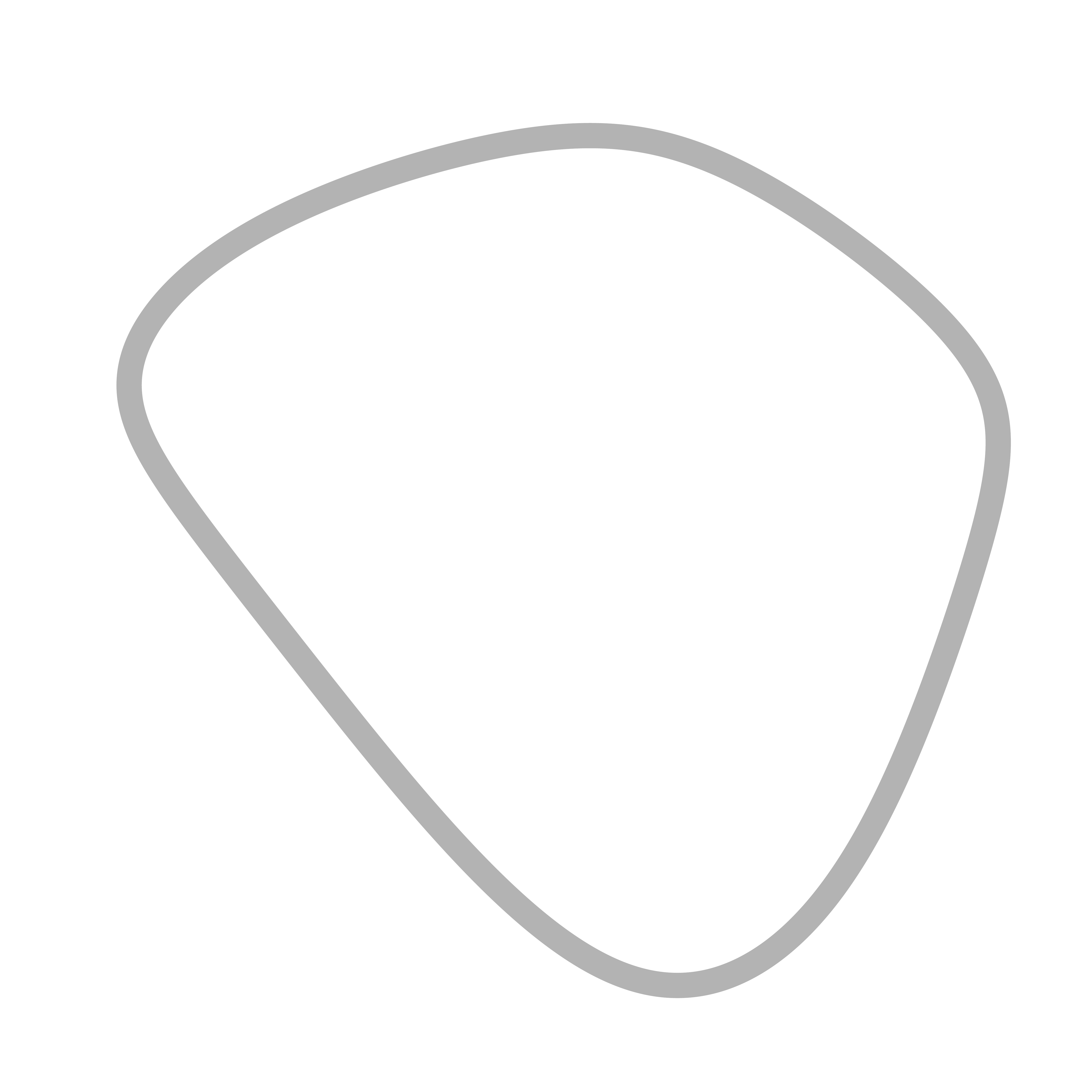}}\hspace{3ex}
	\subfigure[7 samples]{\includegraphics[width=0.12\textwidth]{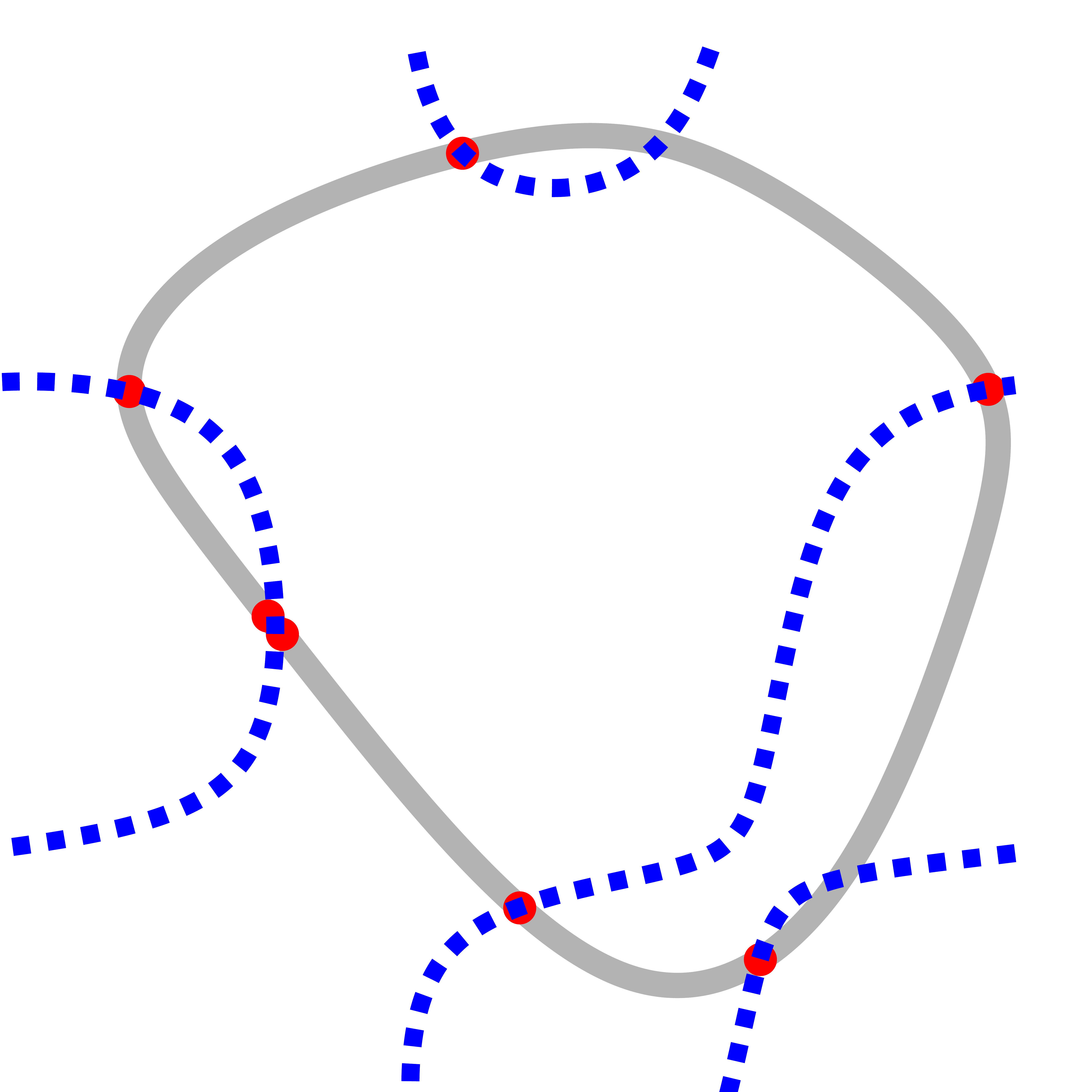}}\hspace{3ex}
	\subfigure[8 samples]{\includegraphics[width=0.12\textwidth]{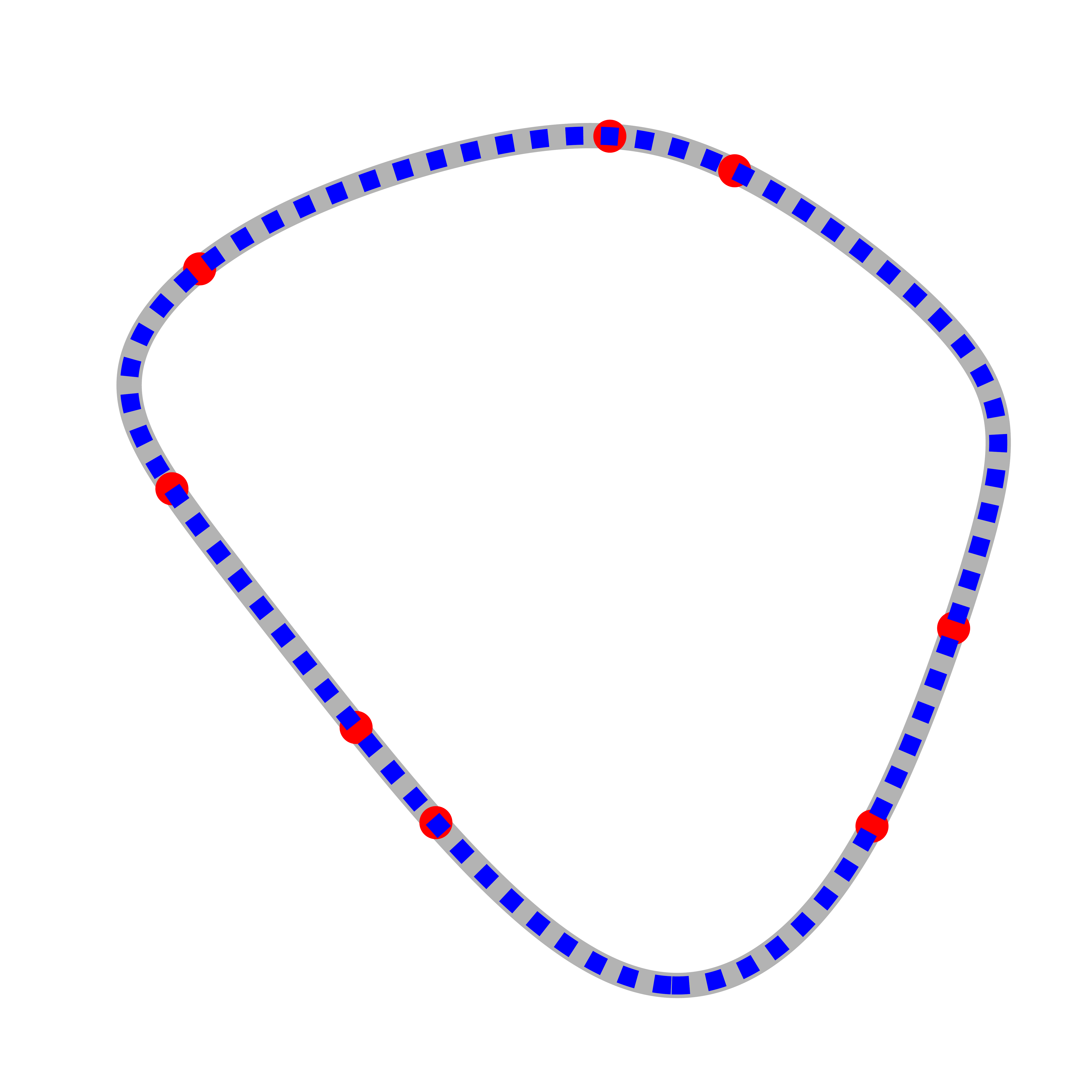}}
	\caption{Illustration of Theorem \ref{rankprop} in the planar setting. The irreducible curve $\mathcal C$, shown in (a) is the zero level-set of a trigonometric polynomial whose bandwidth is $3\times 3$. According to Theorem \ref{rankprop}, we will need at least 8 samples to recover the curve. In (b), we randomly choose 7 samples (the red dots) on the original curve (the gray curve) and the blue dashed curve shows the recovery curve from this 7 samples. Since the sampling condition is not satisfied, the recovery failed. In (c), we randomly choose 8 points (the red dots). From (c), we see that the blue dashed curve (recovery curve) overlaps the gray curve (the original curve), suggesting perfect recovery.}
	\label{illu2d}
\end{figure}
When $n=2$, the surface reduces to a planar curve, which is the case considered in \cite{poddar2018recovery,poddaricassp}. Specifically, if the bandwidth $\Lambda=k_1\times k_2$ is specified by a rectangular region, the results in \cite{poddar2018recovery,poddaricassp} shows that it can be recovered from $(k_1+k_2)^2$ samples. By contrast, the above recovery guarantees reduces the sampling requirement to $|\Lambda|-1=k_1\cdot k_2-1$, which is essentially the degrees of freedom of the curve. This quite significant reduction in the number of samples is obtained by relaxing the recovery conditions from worst case to high probability. Specifically, it is possible to come up with $k_1\cdot k_2-1$ or more samples on $\mathcal C$, such that the rank of the feature matrix is less than $|\Lambda|-1$; however, the probability for such a choice is zero, when the samples are chosen randomly. Note that the gain in sampling is considerably more significant in high dimensional setting, where the direct extension of the results in \cite{poddar2018recovery,poddaricassp} suggests $(\sum_{i=1}^n k_i)^n$ samples, while the proposed approach only requires $(\prod_{i=1}^n k_i) -1$ samples. For example, when $k_i = 5; i=1,2,3$, the worst case guarantee requires $3375$ points, while the high probability guarantee only needs $124$ samples.  

\begin{figure}[t!]
\centering
\subfigure[Surface]{\includegraphics[width=0.15\textwidth]{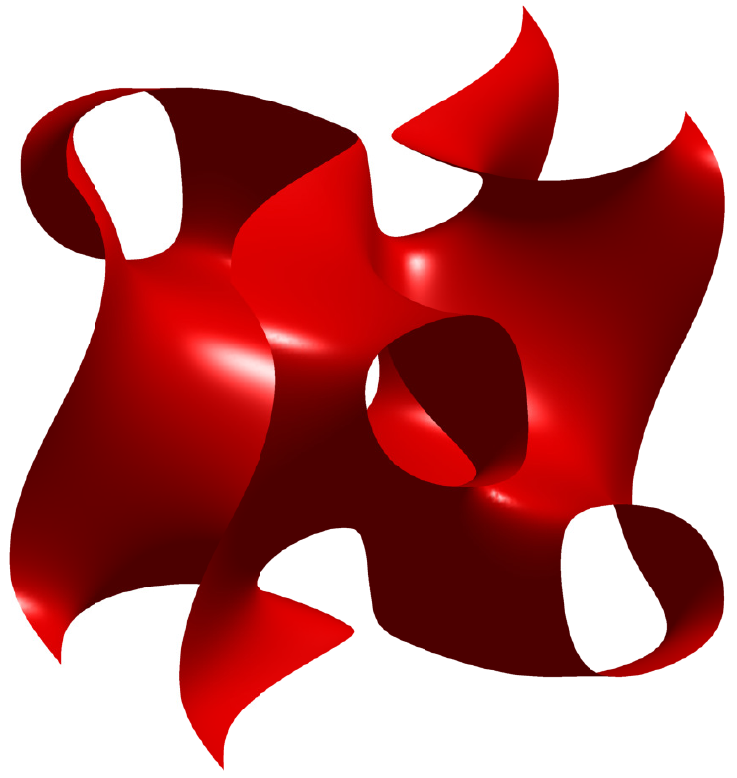}}
\subfigure[25 samples]{\includegraphics[width=0.15\textwidth]{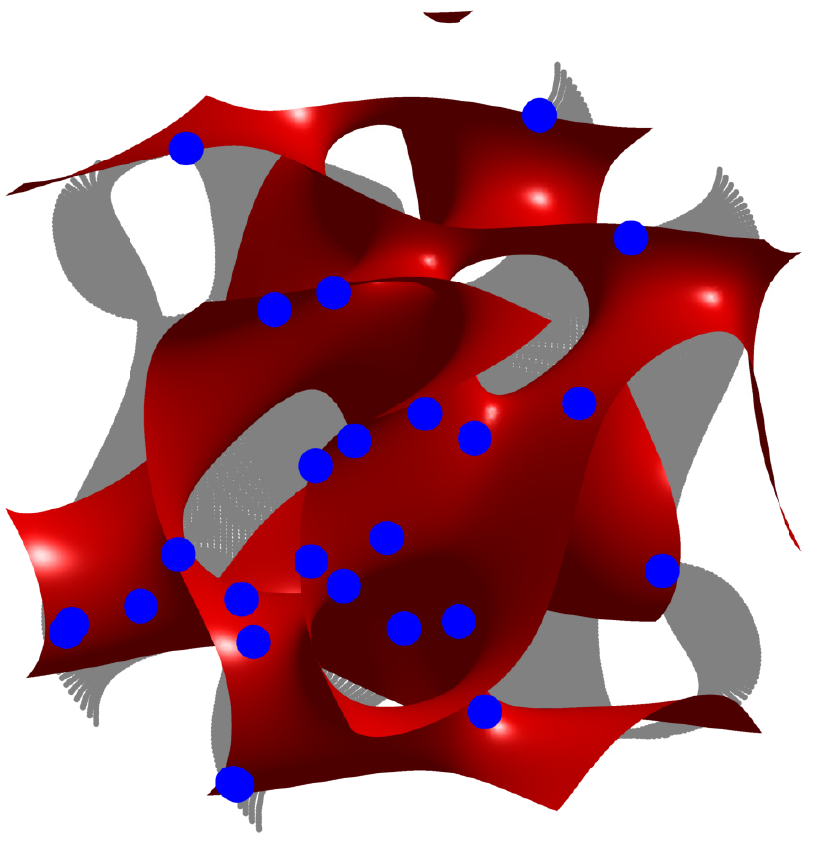}}
\subfigure[26 samples]{\includegraphics[width=0.15\textwidth]{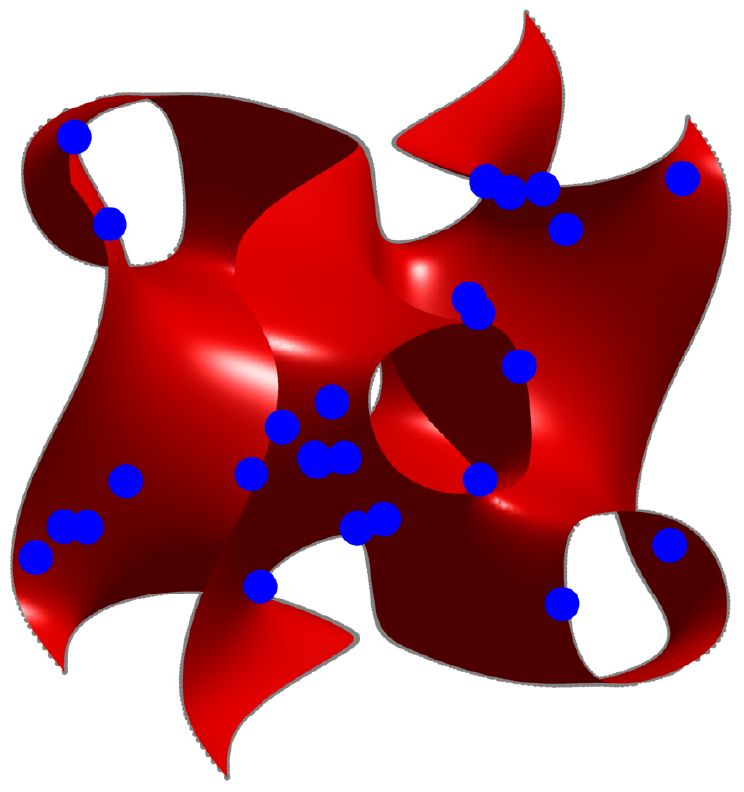}}
\caption{\small Illustration of Theorem \ref{rankprop} in 3-D. The irreducible surface given by (a) is the original surface, which is given by the zero level-set of a trigonometric polynomial whose bandwidth is $3\times 3\times 3$. According to Theorem \ref{rankprop}, we will need at least 26 samples to recover the surface. In (b), we randomly choose 25 samples, indicated by the blue dots on the original surface shown in gray. Note that the recovered surface shown in red differs significantly from the gray one. In (c), we randomly choose 26 points denoted by the blue dots. From (c), we see that the recovered red surface overlaps the gray surface perfectly, suggesting perfect recovery.}\vspace{-1.5em}
\label{illu3d}
\end{figure}

Once the feature matrix $\Phi_{\Lambda}(\mathbf X)$ is constructed with the sufficient number of points, one can uniquely identify the coefficient vector $\mathbf c$ satisfying \eqref{nullspacecond} using eigen decomposition. Here, we assume that the bandwidth $\Lambda$ is perfectly known. The recovered curves and surfaces using
sampling result is illustrated in 2-D in Fig. \ref{illu2d}, while the demonstration in 3-D is shown in Fig. \ref{illu3d}.

In practice, the exact bandwidth of the surface specified by $\Lambda$ is unknown. In this case, we propose to over-estimate the bandwidth as $\Gamma \supset \Lambda$. In this case, \eqref{rankcond} gets modified as
\begin{equation}\label{rankgamma}
\mathrm{rank}(\Phi_{\Lambda}(\mathbf{X}))=|\Gamma|-|\Gamma:\Lambda|,
\end{equation}
where $\Gamma:\Lambda$ denotes the set of all possible shifts of the set $\Lambda$ within $\Gamma$. See \cite{poddar2018recovery,poddaricassp} for details. We do not give a sampling condition for this case in this paper.

\vspace{-1em}
\subsection{Union of irreducible surfaces}
\vspace{-0.5em}
Theorem \ref{rankprop} focused on sampling of an irreducible surface. In practice, one often has composite surfaces, which is the zero level-set of $\psi=\psi_1\cdot \psi_2\cdots \psi_M$, where $\psi_i: i=1,\cdots,M$ are irreducible polynomials. Here, the composite curve/surface is the union of irreducible curves/surfaces, specified by $\mathcal S = \bigcup_{i=1}^M \mathcal S_i$, where $\mathcal S_i$ is the irreducible curve/surface corresponding to $\psi_i$ of bandwidth $\Lambda_i$. The product relation in space domain translates to convolution relation in the Fourier domain, which gives $\mathbf c = \mathbf c_1 * \mathbf c_2*\cdots *\mathbf c_M$. The bandwidth of $\psi$ denoted by $\Lambda$ is thus related to the individual bandwidths $\Lambda_i$. We now consider the recovery of $\mathcal S$ from its samples, which is a non-linear generalization of the results in the context of  of union of subspaces. 

\begin{thm}\label{rankprop2}
Let $\mathcal S = \bigcup_{i=1}^M \mathcal S_i$ be a union of irreducible surfaces, where $\mathcal S_i: i=1,\cdots,M$ is the zero level-set of the irreducible bandlimited function $\psi_i$ of bandwidth $\Lambda_i$. 
Assume that each of the surface $\mathcal{S}_i$ are randomly sampled with $N_i$ points, chosen independently on the zero level-set of $\psi_i$. Then, the surface $\mathcal S$ can be uniquely recovered with high probability iff 
\begin{equation}\label{individual}
N_i \geq  |\Lambda_i|-1; ~i=1,\cdots,M
\end{equation}
and 
\begin{equation}\label{joint}
N = \sum_{i=1}^M N_i \geq  |\Lambda|-1.
\end{equation}
\end{thm}

\begin{figure}[!h]
\centering
\subfigure[Curve]{\includegraphics[width=0.125\textwidth]{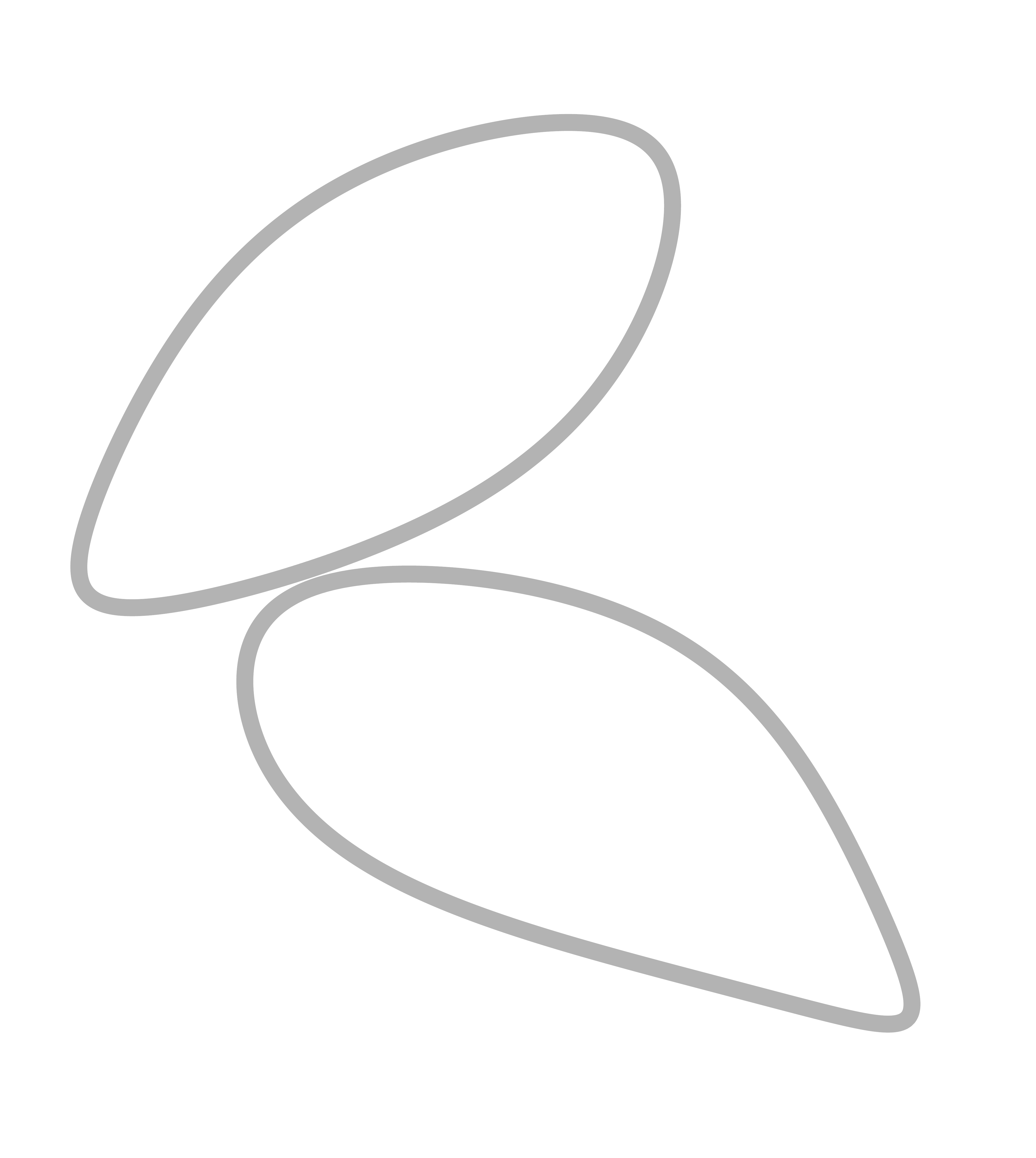}}
\subfigure[7+17]{\includegraphics[width=0.125\textwidth]{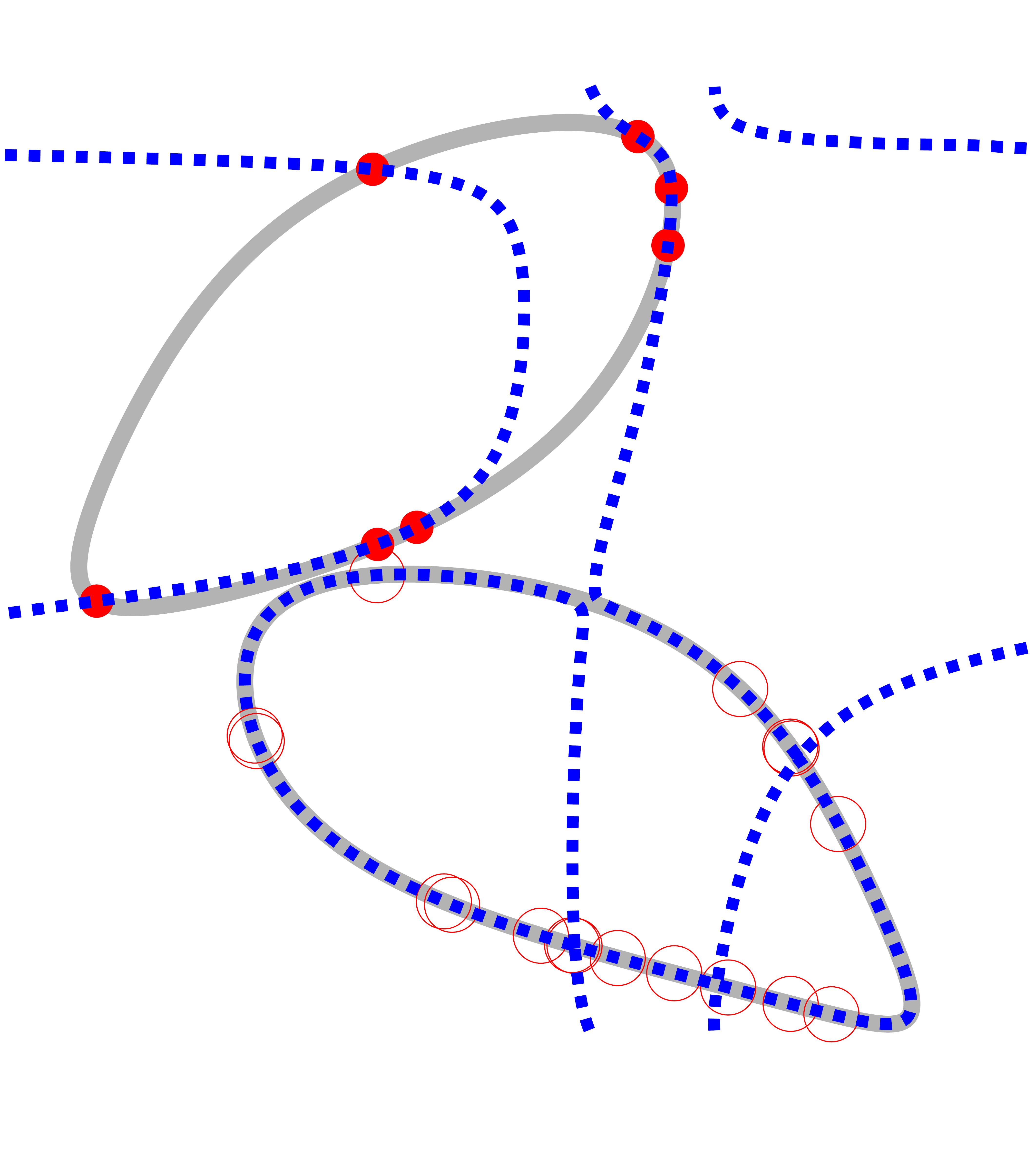}}
\subfigure[8+16]{\includegraphics[width=0.125\textwidth]{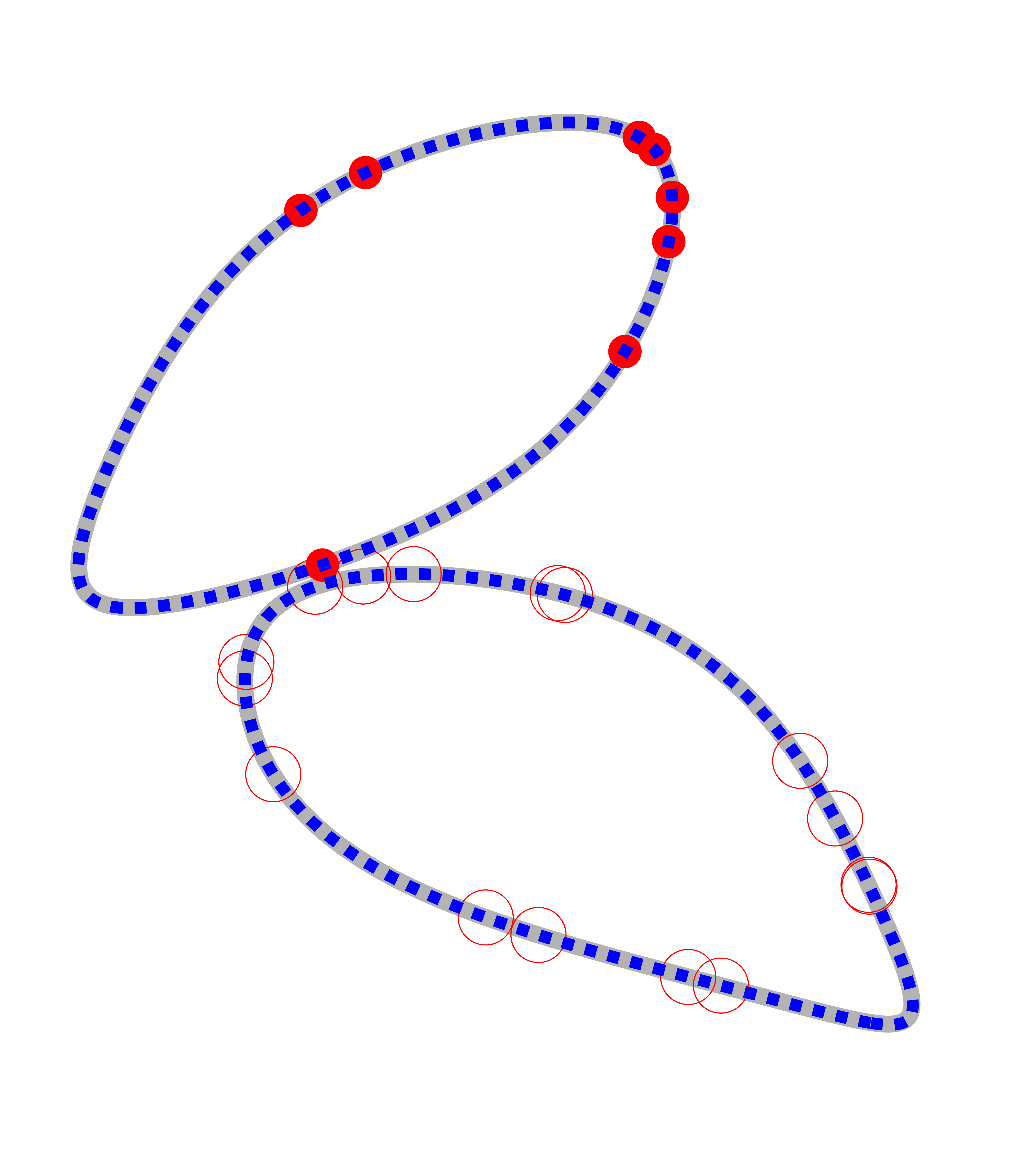}}\\
\subfigure[17+7]{\includegraphics[width=0.125\textwidth]{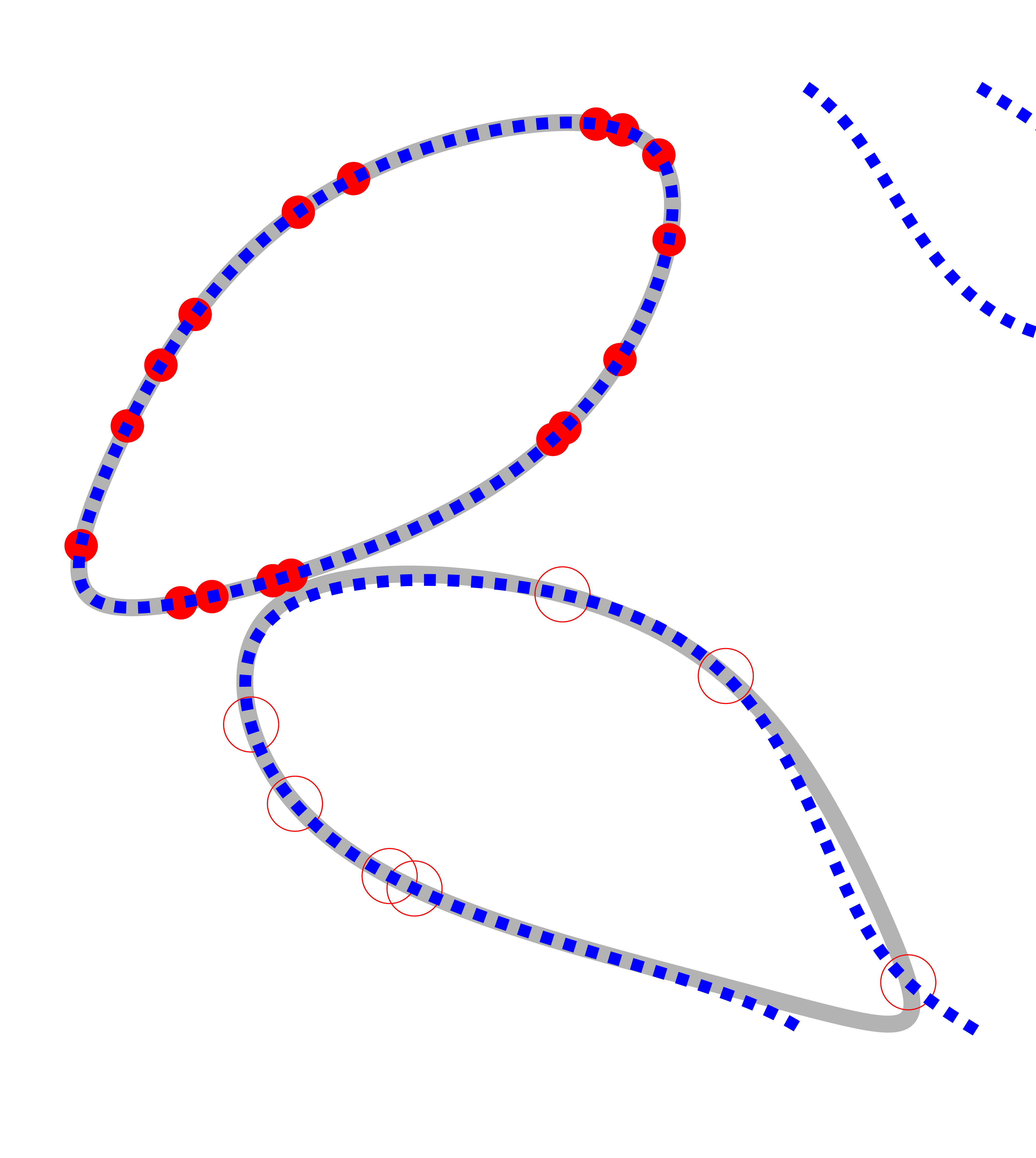}}
\subfigure[16+8]{\includegraphics[width=0.125\textwidth]{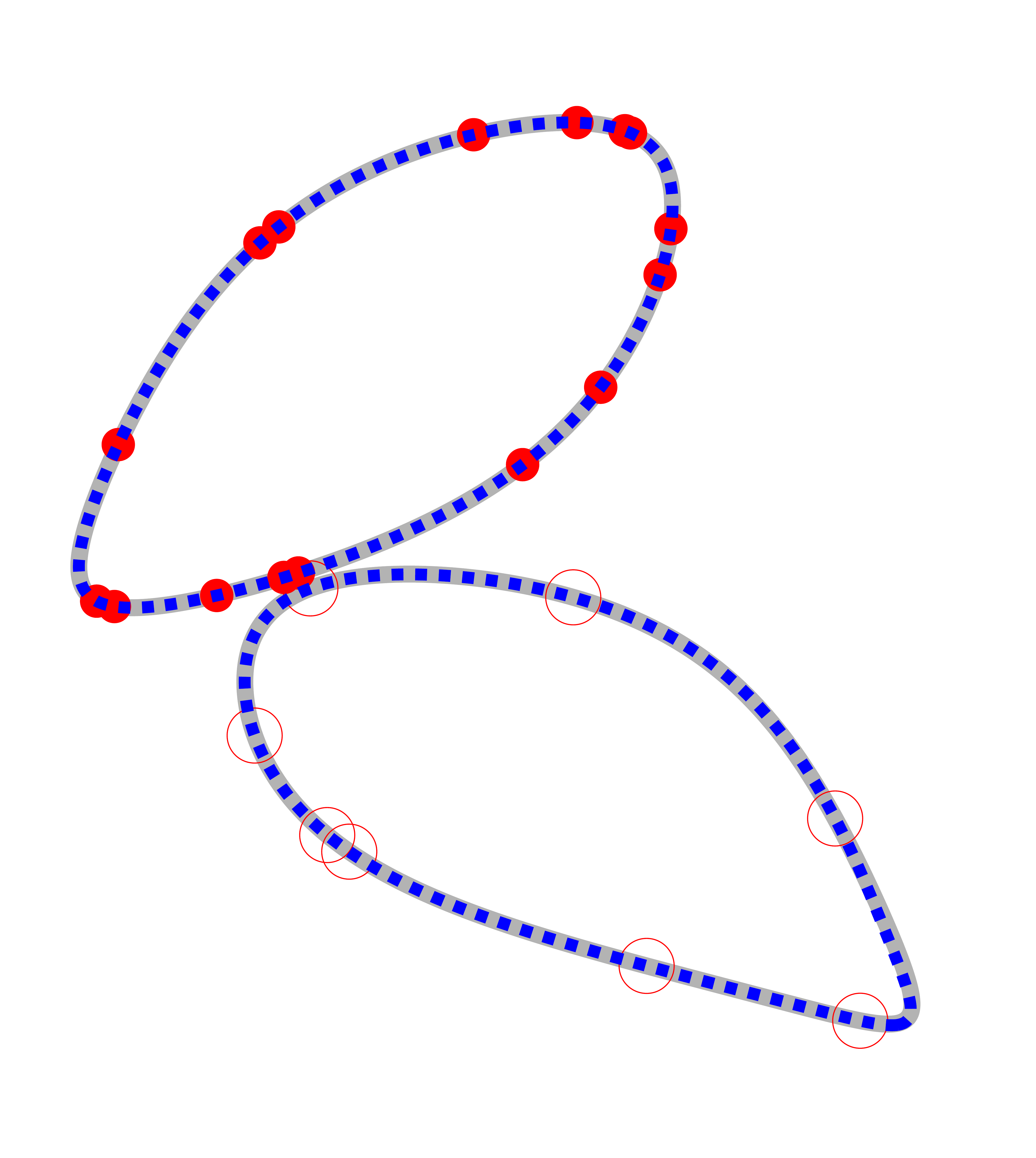}}
\subfigure[8+8]{\includegraphics[width=0.125\textwidth]{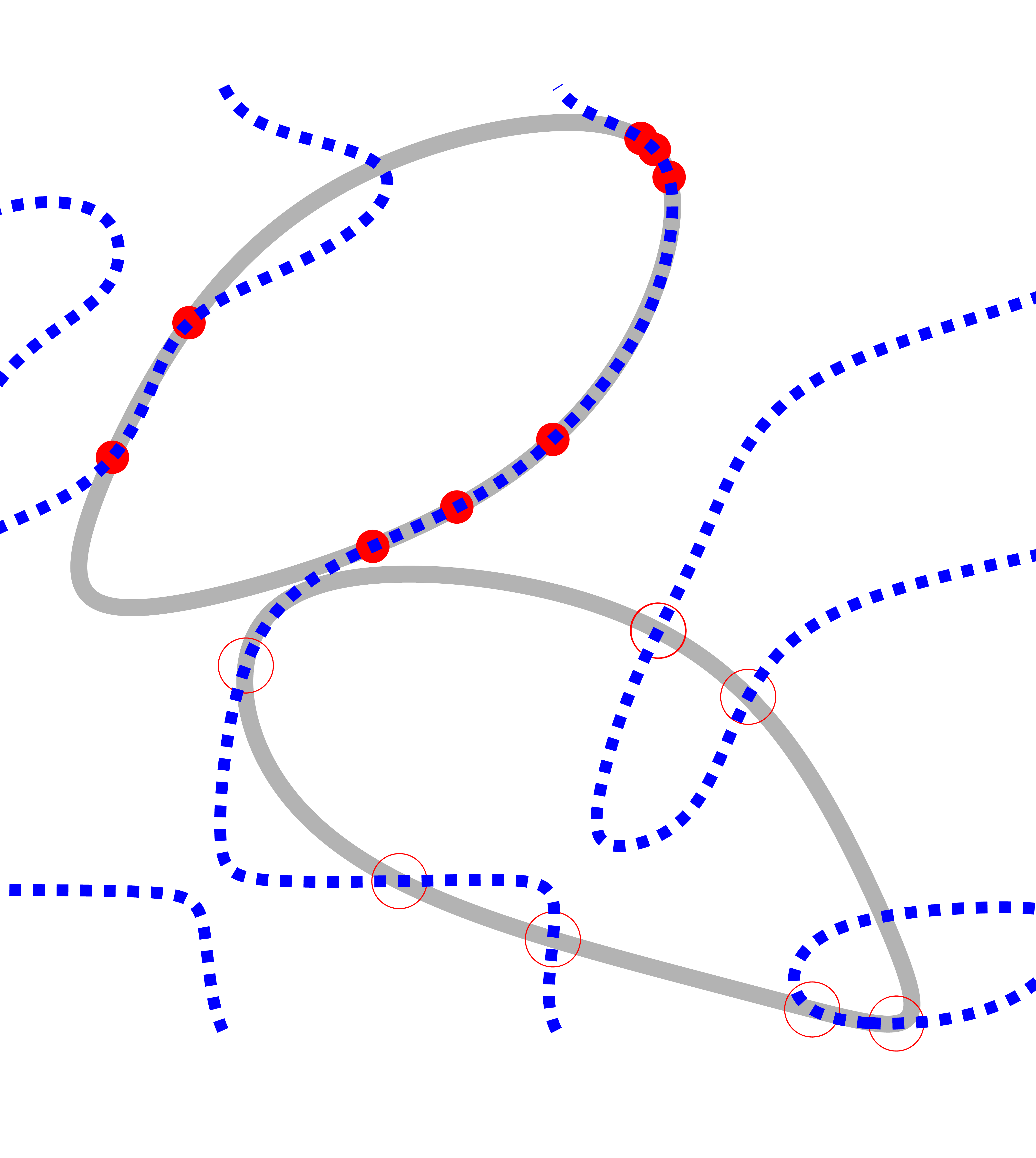}}
\caption{\small Illustration of Theorem \ref{rankprop2}. The original curve (a)  is $\mathcal C = C_1\bigcup C_2$, which is the union of two irreducible curves, each of bandwidth $3\times 3$. According to Theorem \ref{rankprop2}, we totally need at least 24 samples and each of the component need to be sampled at least 8 samples. We consider the recovery of the curve from 24 points, where each irreducible curve is sampled with different number of points as described in the subfigure captions; the first number denotes the number of samples on $\mathcal C_1$, while the second number indicates the number of samples on $\mathcal C_2$. We note that the sampling conditions are satisfied in (c) and (e), when recovery succeeds. By contrast, the recovery fails when any of the components are sampled with fewer points, as shown in  (b) and (d). In (f), we choose 8 samples on each of the component. While the recovery still fails since the total number of samples is not satisfied.}
\label{illuprop2}
\end{figure}

This theorem is true for any dimensions, including the planar setting. Note that unlike the sampling results in Theorem \ref{rankprop}, the samples cannot be randomly chosen on the curve. The number of samples on each irreducible component should be proportional to the complexity of the curve. The interesting observation is that the sampling of each curve proportional to its complexity specified by \eqref{individual} is not alone sufficient for perfect recovery. For example, we consider a planar setting in Fig. \ref{illuprop2}, where the curve is the union of two curves, each of bandwidth $3\times 3$. According to the above result, we require each component to be sampled with a minimum of $|\Lambda_i|-1=8$ points, while the total number of samples should be $|\Lambda|-1=24$, which exceeds $2(|\Lambda_i|-1) = 18$. 

\vspace{-2.3em}
\section{Local representation of functions}
\vspace{-0.9em}
We now consider the efficient representation of complex functions in high dimensional spaces. We note that learning such functions from measured data is a key problem in machine learning applications. The direct representation of such functions suffers from the curse of dimensionality. The large number of parameters needed for such a representation makes it difficult to learn such functions from few labeled data points. 

Fortunately, natural data often lies on simpler constructs such as surfaces in high dimensional space. We now show that a bandlimited multidimensional function can be perfectly represented over a union of surfaces with a fraction of function samples. We will focus on bandlimited multidimensional functions of the form 
\begin{equation}\label{fn}
f(\mathbf {x})=\sum_{\mathbf {k}\in\Gamma} {a_k}\exp(j2\pi \mathbf {k}^T\mathbf {x}) = \mathbf{a}^T~\Phi_{\Gamma}(\mathbf{x})
\end{equation}
Note that the direct representation of the function requires $|\Gamma|$ coefficients, which suffers from the curse of dimensionality. 

We now use \eqref{rankgamma}, which suggests that the rank of the feature matrix will be at most $|\Gamma|-|\Gamma:\Lambda|$, which is far smaller than $|\Gamma|$, to come up with an efficient representation. Note that the rank property specified by \eqref{rankgamma} is valid only when the points are located on $\mathcal S$. Similar to Theorem \ref{rankprop}, we have that if we randomly distribute points on $\mathcal S$, the feature matrix will satisfy \eqref{rankgamma} with high probability. The above results imply that the feature vector for any point  on $\mathcal S$ can be computed as the linear combination of feature vectors $\Phi_{\Gamma}(\mathbf x_i); i=1,..,P$, where $\mathbf x_i$ are  $P=|\Gamma|-|\Gamma:\Lambda|$ points on the curve. Solving for the coefficients and using the "kernel-trick", we obtain the following result.

\begin{figure}[!h]
\centering
\subfigure[Curve]{\includegraphics[width=0.155\textwidth]{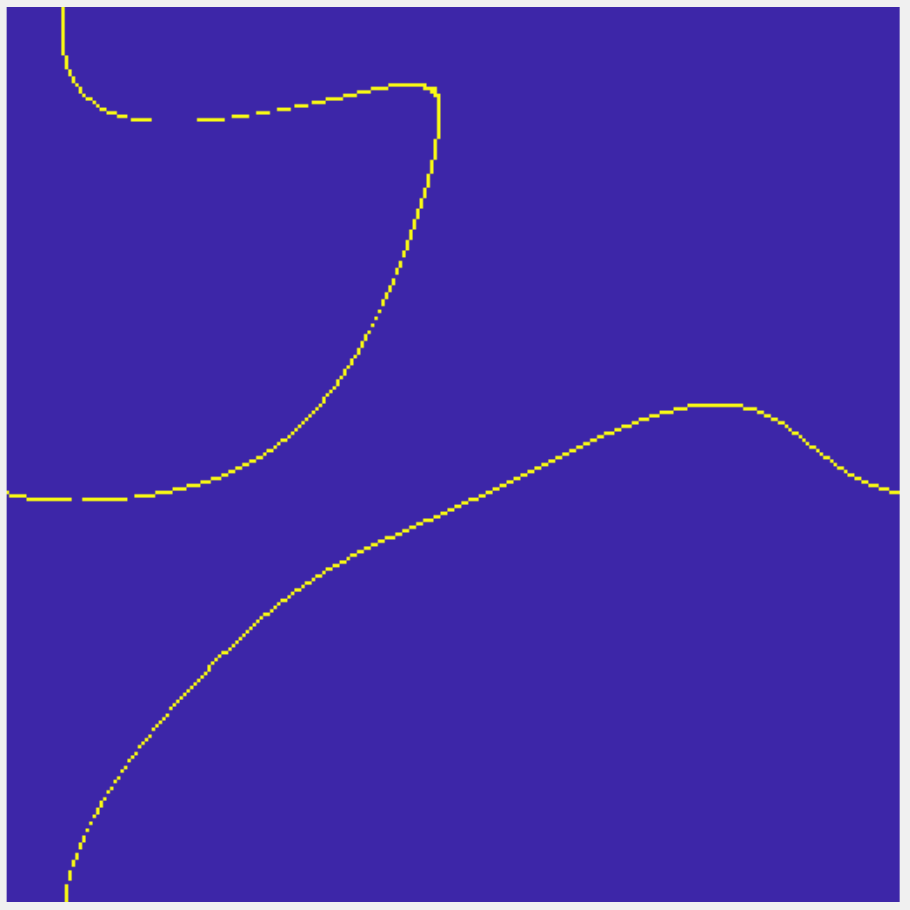}}
\subfigure[BL function]{\includegraphics[width=0.155\textwidth]{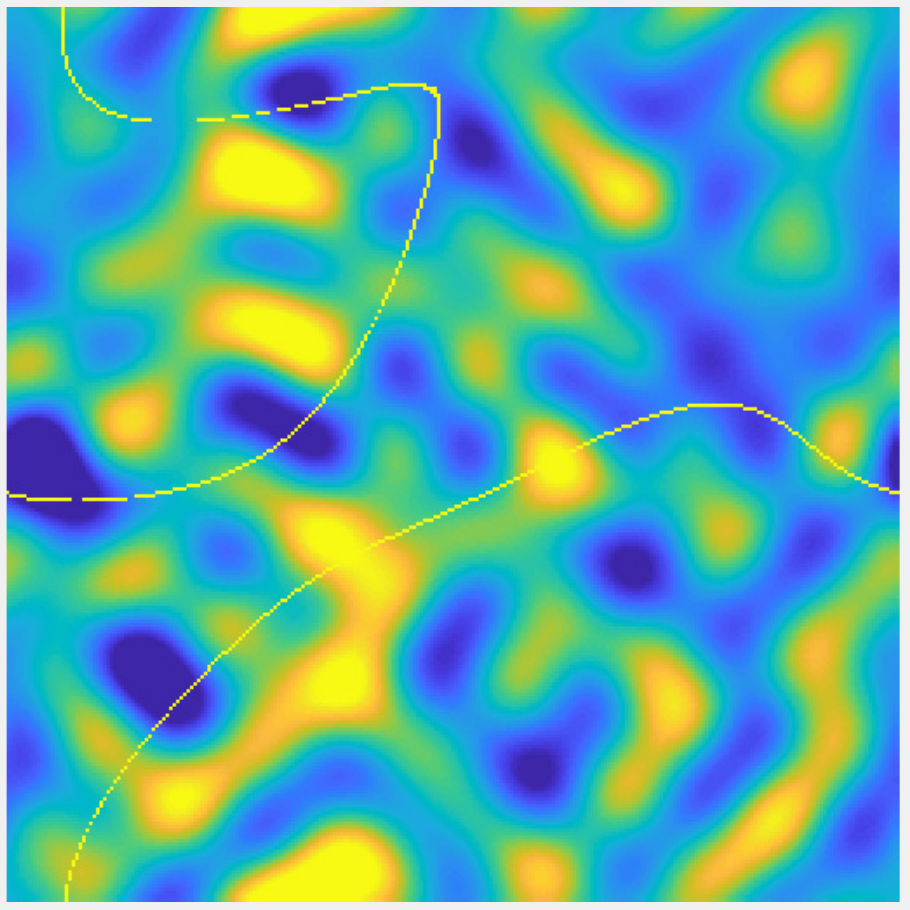}}
\subfigure[Function on curve]{\includegraphics[width=0.155\textwidth]{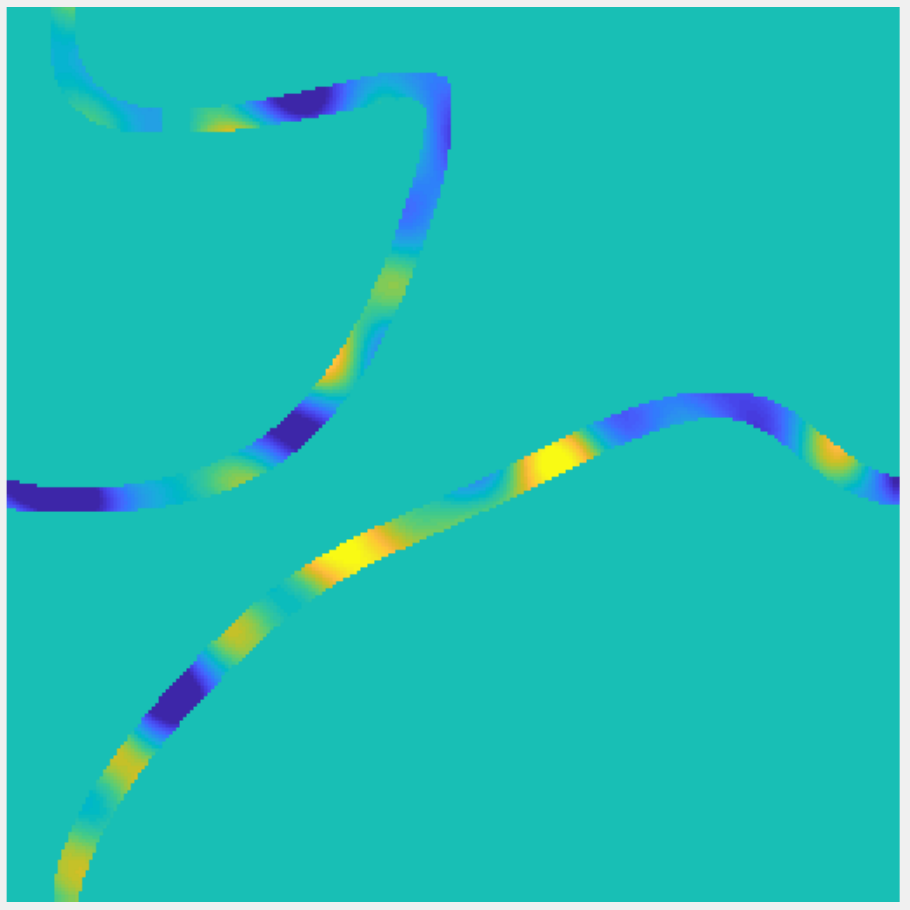}}
\subfigure[Samples]{\includegraphics[width=0.155\textwidth]{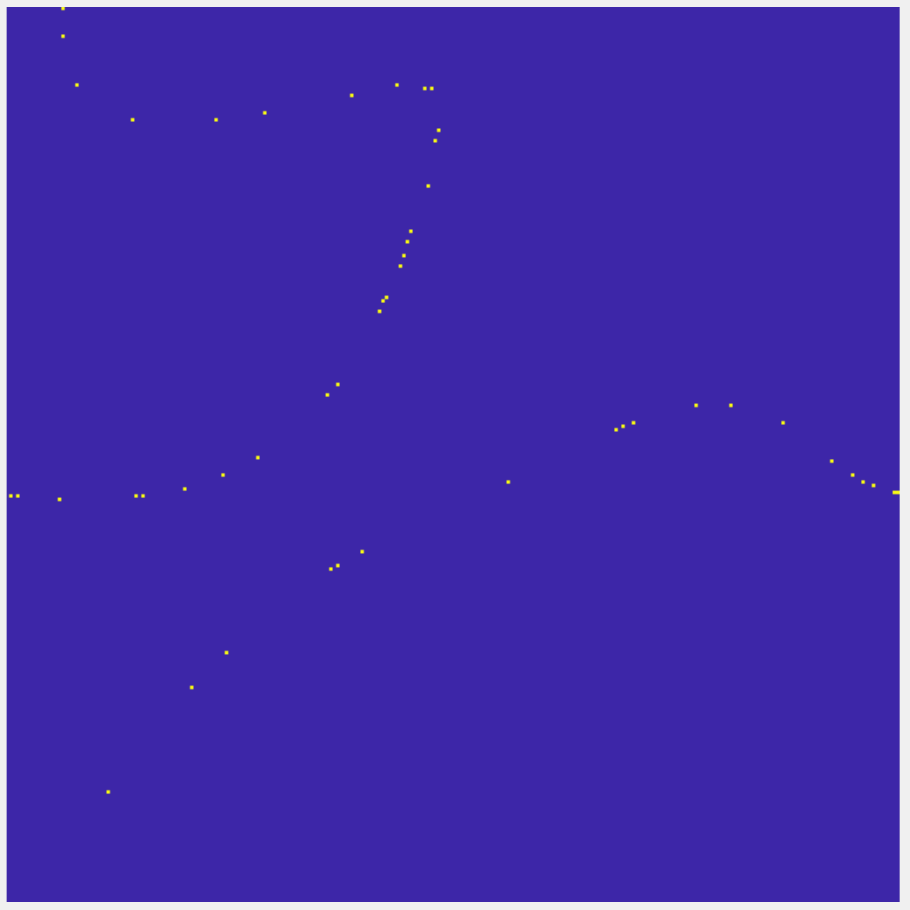}}
\subfigure[Approximation]{\includegraphics[width=0.155\textwidth]{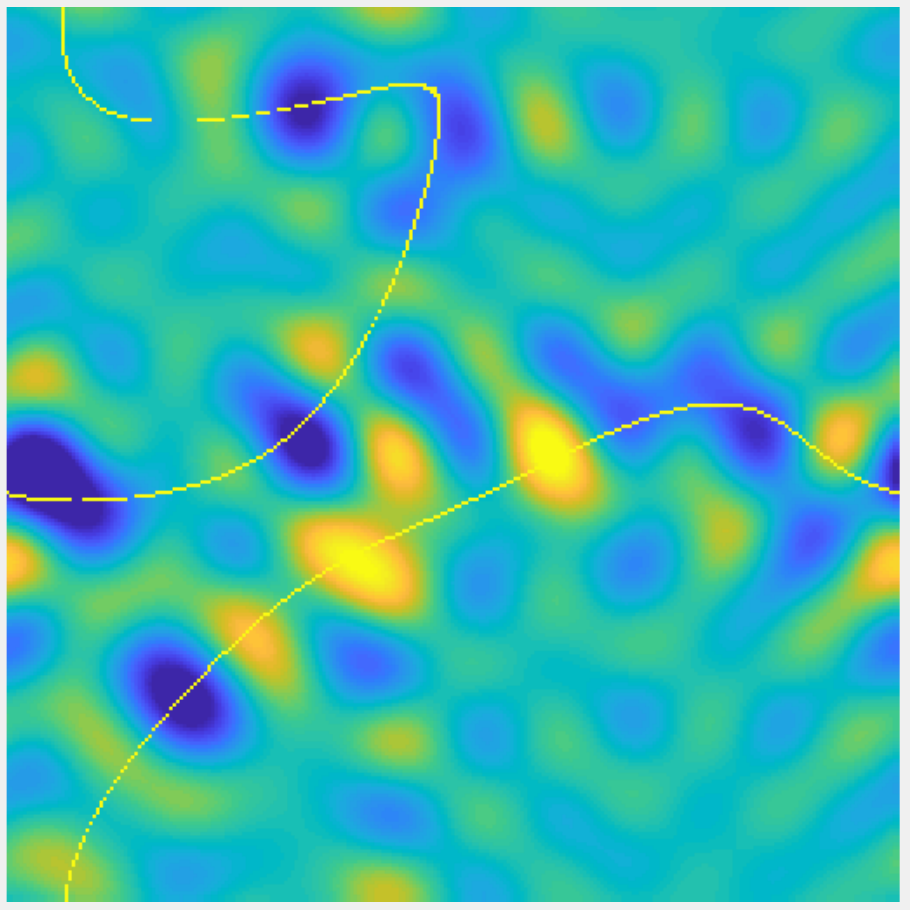}}
\subfigure[Approx on curve]{\includegraphics[width=0.155\textwidth]{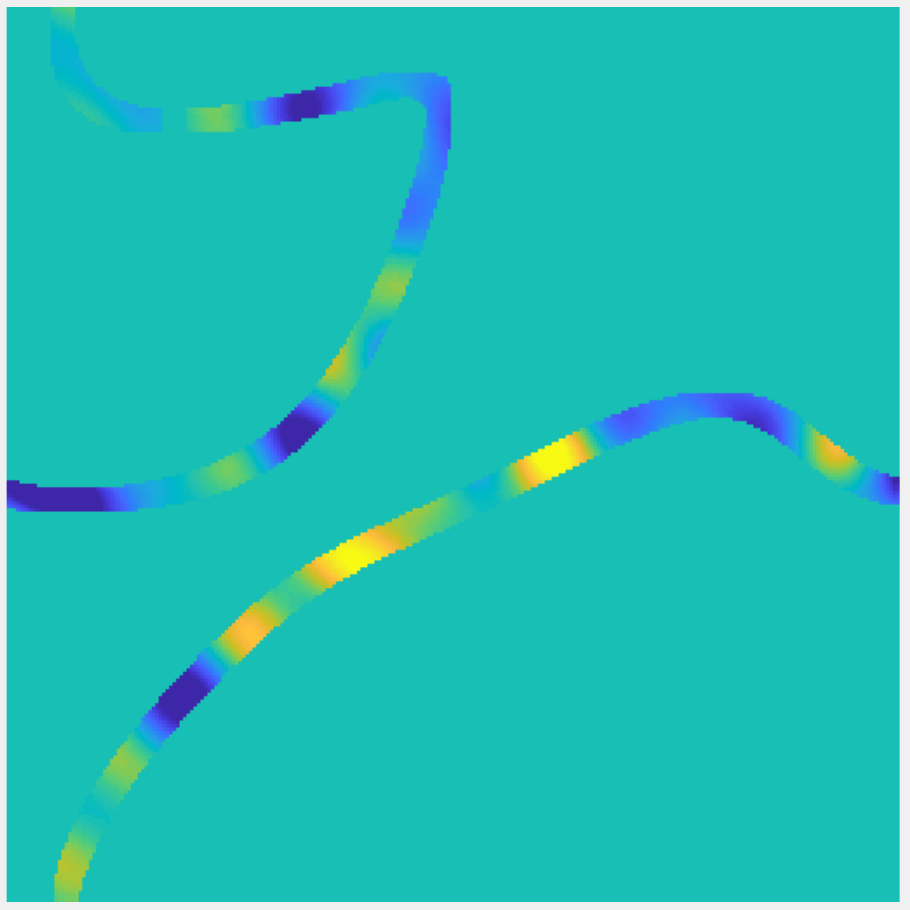}}
\caption{\small Illustration of Proposition \ref{functioncurve} on local representation of functions. We consider the local approximation of the bandlimited function in (a) with a bandwidth of $13\times13$, on the bandlimited curve shown in (b). The bandwidth of the curve is $3\times3$. The curve is overlaid on the function in (a) in yellow color. The restriction of the function to the vicinity of the curve is shown in (c). Our results in Proposition \ref{functioncurve} suggests that the local function approximation requires $13^2 - 11^2 = 48$ anchor points. We randomly select the points on the curve, as shown in (d). The interpolation of the function values at these points yields the global function shown in (e). The restriction of the function to the curve in (f) shows that the approximation is good.} \vspace{-2em}
\label{illus}
\end{figure}

\begin{prop}\label{functioncurve}
\label{functioncurve}
Suppose $\mathcal S$ is an irreducible surface with bandwidth $\Lambda$. Consider an arbitrary bandlimited function specified by $f(\mathbf{x})=\mathbf{a}^T\phi_{\Gamma}(\mathbf{x})$ with bandwidth $\Gamma$ such that $\Gamma \supset \Lambda$. For any arbitrary point on $\mathcal S$, $f(\mathbf x)$ can be exactly represented as \[ \hat{f}(\mathbf{x})=\mathbf{p}^T~\kappa(\mathbf{X}),\]
where $\mathbf{p}^T=\begin{bmatrix} f(\mathbf{x}_1) & f(\mathbf{x}_2) &\cdots & f(\mathbf{x}_{P})\end{bmatrix} \mathbf{K}^{-1}$ and $\kappa(\mathbf{X})=\begin{bmatrix} k(\mathbf{x},\mathbf{x}_1)  & \cdots & k(\mathbf{x},\mathbf{x}_{P})\end{bmatrix}^T$. Here, $\mathbf x_1,\cdots,\mathbf x_P$ are $P=|\Gamma|-|\Gamma:\Lambda|$ points on $\mathcal S$. $k_{\Gamma}(\cdot,\cdot)$ denotes the Dirichlet kernel of bandwidth $\Gamma$ specified by
\[k(\mathbf{x},\mathbf{y})=\phi_{\Lambda}(\mathbf{x})^H\cdot\phi_{\Lambda}(\mathbf{y}).\] 
$\mathbf K$ is the $P\times P$ kernel matrix with entries $\mathbf K_{i,j} = k_{\Gamma}(\mathbf x_i,\mathbf x_j)$.
\end{prop}
The above proposition is illustrated in Fig. \ref{illus} in the 2-D setting. Specifically, the local representation in (e) matches the original function in (b) on the curve. This local representation reduces the number of parameters in the representation from $169$ parameters to $48$ parameters. The reduction in number of free parameters will be even more significant in high dimensions.

\vspace{-1.5em}\section{Conclusion}
\vspace{-0.5em}
In this paper, we considered the recovery of surfaces from few of their samples. We showed that the exponential feature maps of the data points on surfaces lie in low-dimensional subspaces. The low-rank structure of the feature matrix is used to recover the surface from few measurements. Our results show that the surface can be uniquely recovered with high probability if the curves are sampled at a rate higher than the degrees of freedom. These results also provide an efficient approach for the local representation of multidimensional functions on surfaces from few measurements.

\bibliographystyle{IEEEbib}
\bibliography{refs}

\begin{thebibliography}{1}

\bibitem{roweis2000nonlinear}
Sam~T Roweis and Lawrence~K Saul,
\newblock ``Nonlinear dimensionality reduction by locally linear embedding,''
\newblock {\em science}, vol. 290, no. 5500, pp. 2323--2326, 2000.

\bibitem{poddar2018recovery}
Sunrita Poddar and Mathews Jacob,
\newblock ``Recovery of point clouds on surfaces: Application to image
  reconstruction,''
\newblock in {\em Biomedical Imaging (ISBI 2018), 2018 IEEE 15th International
  Symposium on}. IEEE, 2018, pp. 1272--1275.

\bibitem{poddaricassp}
Sunrita Poddar and Mathews Jacob,
\newblock ``Recovery of noisy points on bandlimited surfaces: Kernel methods
  re-explained,''
\newblock in {\em 2018 IEEE International Conference on Acoustics, Speech and
  Signal Processing (ICASSP)}. IEEE, 2018, pp. 4024 -- 4028.

\bibitem{poddar2018sampling}
Sunrita Poddar, Qing Zou, and Mathews Jacob,
\newblock ``Sampling of planar curves: Theory and fast algorithms,''
\newblock {\em arXiv preprint arXiv:1810.11575}, 2018.

\bibitem{ongie2016off}
Greg Ongie and Mathews Jacob,
\newblock ``Off-the-grid recovery of piecewise constant images from few fourier
  samples,''
\newblock {\em SIAM Journal on Imaging Sciences}, vol. 9, no. 3, pp.
  1004--1041, 2016.

\end{thebibliography}

\end{document}